\documentclass[11pt]{article}
\usepackage{epsfig,amscd,amssymb}
\usepackage{amsmath,graphicx}
\newcommand{\be}{\begin{equation}}
\newcommand{\ee}{\end{equation}}
\newcommand{\bea}{\begin{eqnarray}}
\newcommand{\eea}{\end{eqnarray}}

\newcommand{\zn}{{{\mathbb Z}_n}}
\newcommand{\zth}{{{\mathbb Z}_3}}
\newcommand{\ztwo}{{{\mathbb Z}_2}}

\newcommand{\Hone}{{\cal H}^{(1)}}
\newcommand{\Htwo}{{\cal H}^{(2)}}

\newcommand{\ah}{{\widehat{\alpha}}}
\newcommand{\phih}{{\widehat{\phi}}}
\newcommand{\JQ}{{\cal J}^{(Q)}}

\begin{document}

\title{Parafermionic edge zero modes in ${\mathbb Z}_n$-invariant spin chains}
\author{Paul Fendley
\medskip \\ 
Department of Physics, University of Virginia,
Charlottesville, VA 22904-4714 USA\\
}

\smallskip 

\date{October 2, 2012} 

\maketitle

\begin{abstract} 

A sign of topological order in a gapped one-dimensional quantum chain is the existence of edge zero modes. These occur in the $\ztwo$-invariant Ising/Majorana chain, where they can be understood using free-fermion techniques. Here I discuss their presence in spin chains with $\zn$ symmetry, and prove that for appropriate couplings they are exact, even in this strongly interacting system. These modes are naturally expressed in terms of parafermions, generalizations of fermions to the $\zn$ case. I show that parafermionic edge zero modes do not occur in the usual ferromagnetic and antiferromagnetic cases, but rather only when the interactions are chiral, so that spatial-parity and time-reversal symmetries are broken.

\end{abstract} 

\section{Introduction}

One of the fundamental results of statistical mechanics is Onsager's computation of the exact free energy of the two-dimensional Ising model \cite{Onsager1}. This provided a canonical example of a non-trivial phase transition between an ordered phase, where the $\ztwo$ spin-flip symmetry is spontaneously broken, and a disordered phase.
This computation and many others in the Ising model are greatly simplified by Kaufman's rewriting of transfer matrix \cite{Kaufman} in terms of free fermions. These results apply to different  couplings in the two directions,  including the limit where the 1d quantum Ising Hamiltonian can be extracted from the transfer matrix. In this limit, Kaufman's mapping amounts to the Jordan-Wigner transformation of spins to fermions \cite{SML}.  
However, because this mapping is non-local, computing many other quantities of interest requires considerably more effort \cite{McCoyWu}. A gargantuan amount of study therefore has been devoted to this model in the many years since these computations, and as a consequence, many aspects are now very well understood. 

Nonetheless, some fundamental aspects of the Ising model have only been understood well in recent years. In particular, Kitaev gave a dramatically different way of understanding the phases and the physical implications of the model \cite{KitMajorana} in its {\em fermionic} description. The point is that  a ``physical'' quantity is typically measurable locally, so the fermions in the 1d quantum Ising chain are unphysical from the point of view of the spin system; the mapping to fermions is a very useful but purely mathematical trick. However, since electrons can be trapped in one dimension (for example in a ``quantum wire''), a relevant question is: What is the physics of the quantum Ising chain when the fermions are the underlying physical degrees of freedom? 

Reinterpreting the quantum Ising Hamiltonian in fermionic language is very interesting but not very difficult.  Since there is no $U(1)$ symmetry in the Ising chain, the number of fermions is only conserved mod 2; in fermionic language this means there is a Cooper pairing interaction. Working out the fermion dispersion relation shows the unusual feature that there is only one fermi point: as opposed to typical models of fermions on the lattice, there is no doubling. In currently fashionable parlance, the fermion is {\em Majorana}.\footnote{Surprisingly, the word ``Ising'' does not appear in Kitaev's paper (except inside ``Surprisingly'') nor in the excellent review article \cite{Jason}; often this chain is now referred to as the ``Kitaev chain''. In an effort to restore interspecies harmony, I will usually call it the ``Ising/Majorana" chain when written in terms of fermionic variables.} 

Less simple, however, is giving a fermionic interpretation of the phases.  The order parameter is simply the magnetization, local in the spin variables, but non-local in the fermions. In the ordered phase, there are two degenerate ground states, so that the $\ztwo$ spin-flip symmetry exchanging them is spontaneously broken. Kitaev's remarkable observation is that this Ising spin order becomes in fermionic language an example of {\em topological order}. A signal of topological order is the appearance of degenerate ground states {\em without} a local order parameter being spontaneously broken. The Ising chain rewritten in terms of Majorana fermions fits the bill; the local order parameter in terms of spins becomes non-local. 

Taking open boundary conditions is an excellent way of probing the situation. Gapless or zero-energy {\em edge modes} often (although not always) occur in a phase with topological order. In the Ising/Majorana chain, the ${\mathbb Z}_2$ conserved quantity is $(-1)^F$, the fermion number mod 2. One of the two ground states thus can be taken to have $(-1)^F=1$, the other $(-1)^F=-1$. Any operator $\Psi$ mapping one such ground state to the other must necessarily be fermionic, i.e.\ obey $(-1)^F\Psi = - \Psi(-1)^F$. A fermionic zero mode, however, does more than just map between the ground states: it commutes with the Hamiltonian, so that the {\em entire spectrum} in the two sectors $(-1)^F=\pm 1$ must be the same. Since the phase with spin/topological order is gapped, such a zero mode can be localized at the {\em edge} of the system. For the Ising/Majorana chain, there are two such exact fermionic edge zero modes \cite{KitMajorana}. These undoubled (or ``unpaired'') Majorana edge zero modes are a defining characteristic of topological order in this system.

This observation has attracted a great deal of attention because in the gapped topological phase, the two ground states can be used to make a ``qubit'', a quantum two-state system robust against decoherence. It is robust because the gap makes transitions to any excited states essentially impossible, while transitions between the two ground states can only be caused by processes that add or remove an electron from the system. Acting with both zero modes does not change $(-1)^F$, but since the modes are localized at opposite ends of the system, no local noise can cause this change. The non-locality of the order parameter in the fermionic basis is thus a feature, resulting in this protection against decoherence. Various proposals have been made to realize the Ising/Majorana physics in real fermion systems, and there are reports of observation \cite{Jason}.

A natural next step is to understand the effects of including interactions between the fermions. In this one-dimensional system or a collection of coupled such systems, the answer is known. As long as the number of chains is non-zero mod 8, the only way to go from topological order to another phase is to close the gap, i.e.\ tune the parameters through a quantum critical point \cite{Fidkowski}. The topological order that remains is essentially the same as that of the systems without interactions. Finding topological order in 1d systems that are ``far'' from free fermions (i.e.\ those that can not be obtained by deformation) is a much more complicated question. A general classification scheme exists \cite{ChenGuWen,Turner,Schuch}, but  connecting this to simple models is non-trivial. 

It thus seems a good idea to go back to spin systems to understand if any simple-to-describe models also can be reinterpreted in terms of topological order. A natural set of candidates are {\em clock} models, where the two-state Ising spin variable and a $\ztwo$ symmetry of Ising are generalized to having $n$ states and a $\zn$ symmetry. Not only do these models exhibit continuous phase transitions between order and disorder \cite{TL}, but they exhibit an even richer spectrum of behavior in the presences of {\em chiral} interactions, where spatial-parity and time-reversal symmetries are broken \cite{Ostlund,Huse,Howes}.  Even more enticingly for present purposes, the clock models can be rewritten in terms of {\em parafermions}, $\zn$ generalizations of fermions \cite{FK}.

The purpose of this paper is to describe how edge zero modes can occur in spin chains with $\zn$ symmetry. These modes are described naturally in terms of the parafermions.  For certain choices of couplings in these systems, I prove that {\em exact} edge zero modes exist, analogous to the Ising/Majorana chain. This proof applies even to spatially varying couplings, i.e.\ the ``random'' case. The corresponding phases thus are presumably topologically ordered. 

One result of the analysis here is that the most widely-studied examples of clock models, those with ferromagnetic or antiferromagnetic interactions, do not have edge zero modes (at least by this construction). Rather, edge zero modes generalizing the Majorana ones only occur if spatial-parity and time-reversal symmetries are broken, i.e.\ the model is chiral. This may seem peculiar, but in the classification schemes for topological order in free-fermion systems in arbitrary dimensions,  discrete symmetries under  time reversal symmetry charge conjugation play central roles  \cite{KitK,Ryu}. Breaking or including these discrete symmetries can indeed change the type of topological order, or eliminate it altogether. The same principle seems to be applicable to this strongly interacting system.

As will be described below, chiral interactions in the clock models result simply from including phases in the nearest-neighbor interaction terms; in fact, one can continuously interpolate between ferromagnetic and antiferromagnetic interactions  by changing these phases.  The most robust type of interaction for topological order seems to be essentially halfway in-between ferromagnetic and antiferromagnetic. In this case the model is {\em symmetric}, having a discrete ``charge-conjugation" type operation that takes the Hamiltonian $H\to -H$ while leaving the spectrum invariant, just like the Ising/Majorana chain does.

This analysis here is considerably more involved than in the Ising/Majorana case. The reason is that the parafermions are not free in any sense; one can for example not use Wick's theorem to simplify the computation of correlators. Thus while they are exceptionally useful for deriving the zero modes here, they do not automatically make computations as easy in general. In fact, for the nearest-neighbor interactions considered, the chiral clock chain is only integrable in a two-parameter subspace of the full model \cite{Gehlen,AMPTY}. This subspace does not even include the ferromagnetic or antiferromagnetic cases, except at the critical points \cite{FZ1}. Rewriting the model in terms of parafermions does however shed light on the integrable cases, in fact giving a simple way of characterizing the integrable cases; I will return to this in a separate publication \cite{me}. 

The outline of the paper is as follows. In section \ref{sec:Ising}, I review some basic facts about the Ising model, including the presence of fermionic edge zero modes in the ordered phase. The $\zn$-invariant spin chains and their parafermionic descriptions are introduced in section \ref{sec:zn}. The topic of section \ref{sec:edge} is the parafermionic edge zero modes, and developing an iterative procedure to find them. They are easily found in an extreme case of the couplings, but even at next order they only occur for chiral interactions. In section \ref{sec:proof}, I prove that for appropriate parameters, the edge zero mode remains exact to all orders. Finally, in section \ref{sec:conclusion} I give the conclusions and some ideas for further work. A proof that the number of zero modes (not necessarily edge modes) increases exponentially with the size of the system is given in an appendix.

\section{Zero and shift modes in the Ising/Majorana chain}
\label{sec:Ising}

To begin, I review the quantum Ising chain, following \cite{SML}, and how to detect the topological order there, following \cite{KitMajorana}. 

\subsection{The Hamiltonian and the fermions}

The Hilbert space for the quantum Ising chain consists of a two-state quantum system, i.e.\ a ``spin'' with spin 1/2, at each of the $L$ sites. 
The Hamiltonian is comprised of two types of terms, those that flip a spin at  a given site, and those that give an interaction energy to adjacent spins. Precisely, with open boundary conditions the Hamiltonian is
\be
H_{\rm IM} = - f \sum_{j=1}^L \sigma^x_j - J \sum_{j=1}^{L-1} \sigma^z_j\sigma^z_{j+1}
\label{HIsing}
\ee
where the Pauli matrices $\sigma^a_i$ act non-trivially at each site $i$, and $f$ and $J$ are non-negative and real.
This Hamiltonian has a ${\mathbb Z}_2$ symmetry under flipping all the spins. For reasons to be apparent shortly, it is natural to name this spin-flip operator $(-1)^F$, so that
\be
(-1)^F = \prod_{j=1}^L \sigma^x_j \ .
\label{min1F}
\ee
This operator indeed squares to $1$, and it is easy to check that 
$[(-1)^F,H_{\rm IM}] = 0$. Another interesting property of the Ising Hamiltonian is that its spectrum is invariant under sending $H_{\rm IM}\to -H_{\rm IM}$. This follows from the canonical transformation $\sigma^x_j\to -\sigma_j^x$ on all sites, and $\sigma^z_{2k}\to -\sigma^z_{2k}$ on every other site, which preserves the algebra of $\sigma^x$ and $\sigma^z$.  

The physics of the Ising chain is well understood. There are two phases, with a critical point at $f=J$.  The phase for $f<J$ is {ordered}, i.e.\ the two-point function $\langle \sigma_j \sigma_k\rangle$ in the ground state goes to a constant value in the limit $|j-k|$ large. This is obvious in the extreme case $f=0$, where there are two ground states: all spins up, and all spins down.  In the disordered phase $f>J$, there is a unique ground state. This is easily seen in the extreme case $J=0$, where the sites are independent so that the ground state is the eigenstate of $\sigma^x_j$ with eigenvalue 1 on each site.


A classic result of statistical mechanics is that the quantum Ising chain can be mapped onto a model of free fermions \cite{SML}. 
The fermionic operators are given by a non-local combination of the spin operators known as a Jordan-Wigner transformation. This is possible because of the duality of the model; the fermions are defined by multiplying the spin operator by its dual, the disorder operator. Duality is a non-local transformation, so the disorder operator is necessarily non-local in terms of the spin operators. Thus the fermion is non-local in terms of the spins, and vice-versa. Precisely, 
at each site of the chain, there are two ``Majorana'' fermion operators defined by
\begin{eqnarray}
a_j &=& \left(\prod_{k=1}^{j-1} \sigma^x_k\right) \sigma^z_j \ ,\\
\label{adef}
b_j &=&ia_j\sigma^x_j= i \left(\prod_{k=1}^{j-1}\sigma^x_k\right)  \sigma^z_j\sigma^x_j\ .
\label{bdef}
\end{eqnarray}
Because of the non-local ``strings'' attached, operators at different points no longer commute. Instead, they anticommute: 
\begin{equation}
\{ a_j, a_k\} = \{b_j,b_k\} = 
2\delta_{jk},\qquad \{a_j,b_k\} = 0
\label{abcomm}
\ee
for all $j$ and $k$. Note also that that $a_j$ and $b_j$ are hermitian and that each squares to 1. 
It can be useful to define the complex fermions $c^{\dagger}_j= a_j+ib_j$ obeying the usual anticommutation relations.

The open-chain Hamiltonian (\ref{HIsing}) rewritten in terms of these fermionic variables is 
\be
H_{\rm IM} = if \sum_{j=1}^L a_j b_j \ +\ i J \sum_{j=1}^{L-1} b_j a_{j+1} \ .
\label{HIF}
\ee
The  Ising Hamiltonian is easy to rewrite in terms of the complex fermions. One finds it includes ``Cooper-pairing'' terms involving $c_j c_{j+1}$ and $c^\dagger_j c^\dagger_{j+1}$. These terms  do {\em not} conserve fermion number generated by $F=\sum_j c^\dagger_j c_j$, but only fermion number mod 2.
The corresponding symmetry generator $(-1)^F$ is the product of all the Majorana fermion operators:
\be
(-1)^F = \prod_{j=1}^L (-i a_j b_j)\ .
\ee
It commutes with any product of an even number of fermion operators, while it anticommutes with any product of an odd number. For this reason, it measures the number of fermion operators mod 2, hence the name. This symmetry operator is also sometimes called ``fermionic parity'', or simply ``parity'' for short; the latter I avoid to prevent confusion with spatial parity symmetry. 
Since $H_{\rm IM}$ is comprised of fermion bilinears it indeed commutes with $(-1)^F$, and so both operators can be simultaneously diagonalized. The states therefore can be divided into two sectors, even and odd under this ${\mathbb Z}_2$ symmetry. 


\subsection{Edge zero modes}
\label{sec:edgeIsing}

The fermionic formulation makes it simple to find the zero-energy edge modes characteristic of topological order. A {\em fermionic zero mode} $\Psi$ is an {\em operator} that
\begin{itemize}
\item commutes with the Hamiltonian:\quad $[H,\Psi]=0$
\item anticommutes with $(-1)^F$: \quad $\{ (-1)^F,\, H \}=0$
\item has finite ``normalization''  even in the $L\to\infty$ limit:\quad $\Psi^\dagger\Psi =1$.
\end{itemize}
The second property guarantees that $\Psi$ maps the even sector to the odd, and the other two then require that the spectrum in the even and odd sectors be identical. An {\em edge zero mode} has the additional property that it is {\em localized} near the edge. This means that the dependence of matrix elements of $\Psi$ on the fermions in the states a distance $l$ from the edge must be exponentially small in $l$.

It is easy to see that an edge zero mode exists in the extreme case $f=0$ of the Ising chain, where the flip terms vanish. In this case, the fermions $a_1$ and $b_L$ do not appear in the Hamiltonian with open boundary conditions. They therefore commute with all the bilinears in the Hamiltonian, and so
\[ [H_{\rm IM} (f=0),a_1] =  [H_{\rm IM} (f=0),b_L] = 0\ .\]
Since $a_1$ and $b_L$ anticommute with $(-1)^F$ and square to 1, each satisfies the other two properties as well. Thus each is a zero mode, and obviously an edge zero mode as well. 

For the Ising model, there are various formal arguments showing that the edge zero modes persist throughout the ordered phase, and so the spin-ordered phase is the same as the topologically ordered one \cite{KitMajorana}. However, as also noted there, because the fermions are free, one simply can find the edge zero modes exactly. The easiest way to find them is to utilize an iteration procedure. Commuting $H_{\rm IM}$ with $a_1$ gives $-2if b_1$. This result can be written as a commutator by using $ 2b_1 = [b_1a_2,a_2] $, Thus a combination of $a_1$ and $a_2$ commutes with the first two terms in the Hamiltonian: 
$$\left[ if a_1 b_1+ iJ b_1a_2\,,\  a_1+\frac{f}{J} a_2\right]\ =\ 0 \ .$$
This process can be iterated by using
\begin{eqnarray}
\label{Hacomm}
[H_{\rm IM}, a_j] &=& - 2if b_j\ +\ 2iJ b_{j-1}\ , \\
\label{Hbcomm}
[H_{\rm IM}, b_j] &=&  2if a_j\ -\ 2iJ a_{j+1}\ , 
\end{eqnarray}
where by convention $b_0 = a_{L+1} =0$. 
Thus the (almost) zero modes localized around the left and right edges for $f<J$ are given by
\begin{equation}
\Psi_{\rm left} = a_1 + \frac{f}{J} a_2 + \left(\frac{f}{J}\right)^2 a_3 +\dots\ ,\qquad
\Psi_{\rm right} = b_L + \frac{f}{J} b_{L-1} + \left(\frac{f}{J}\right)^2 b_{L-2} +\dots
\end{equation}
These do not quite commute with $H_{\rm IM}$, but instead:
\[ 
[H_{\rm IM},\Psi_{\rm left}] = f\left(\frac{f}{J}\right)^{L-1} b_L,\qquad 
[H_{\rm IM},\Psi_{\rm right}] = f\left(\frac{f}{J}\right)^{L-1} a_1,\
\]
The coefficient is exponentially small in $L$, so in the $L\to\infty$ limit, each commutes with $H_{\rm IM}$. 
Each remains normalizable for $f<J$, so throughout the ordered phase, each is an edge zero mode. Thus Ising spin order indeed translates in the fermionic language to topological order.

An important fact is that the edge zero mode survives in the presence of  couplings varying over space. Namely, for
\be
H= i \sum_{j=1}^L f_j a_j b_j \ +\ i \sum_{j=1}^{L-1} J_j b_j a_{j+1} \ .
\label{Hrandom}
\ee
the left edge mode is modified to 
\[
\Psi_{\rm left} = a_1 + \frac{f_1}{J_1} a_2 + \frac{f_1f_2}{J_1J_2} a_3 +\dots\ ,\\
\]
and analogously for the right. 
As $L\to\infty$, these commute with the more general Hamiltonian. They remain normalizable throughout the ordered phase.
Thus this topological order survives even in the presence of disordered couplings.

\section{Parafermions in the chiral ${\mathbb Z}_n$ chains}
\label{sec:zn}

In this section I introduce ${\mathbb Z}_n$-invariant spin chains and show how to write their Hamiltonian in terms of parafermions, $\zn$ generalizations of fermions.  

\subsection{The chiral clock/Potts model}
\label{sec:classicalPotts}

The most famous generalization of the two-dimensional classical Ising model is the $Q$-state Potts model. The two-state Ising spin is replaced with a ``spin'' with $Q$ states, and the Hamiltonian is invariant under the permutation group $S_Q$, generalizing the Ising $\ztwo$ symmetry. With this symmetry, the only possible nearest-neighbor interaction depends on whether the two spins are the same or different. While this model has the virtue of simplicity, a major disadvantage is that for $Q>4$, the phase transition between ordered and disordered phases is not continuous \cite{TL}. 

It is thus interesting to instead consider a more general model, the ``clock'' model. Here each ``spin'' takes on $n$ values, as with the Potts model, but the interactions are required only to be invariant under $\zn$ symmetry. It is convenient to label the values of the spin $s_j$ at site $j$ by $1, \omega, \omega^2\dots \omega^{n-1}$, where $\omega = e^{2\pi i/n}$. The most general $\zn$-invariant coupling between two spins is then
\be
 -J \sum_{m=1}^{n-1} \alpha_m (s^{*}_j s_k)^m\ .
\ee
with $J>0$ by convention.  To make the energy real, the coefficients must obey $\alpha^*_m = \alpha_{n-m}$. The $\zn$ symmetry comes from sending $s_j\to \omega s_j$ for all $j$. The $S_n$ symmetric Potts case corresponds to having all $\alpha_m$ the same.

A traditional ferromagnetic interaction corresponds to real $\alpha_m$ and $J\alpha_m>0$. 
However, allowing $\alpha_m$ to be complex results in interesting behavior not possible in the Ising case. To understand this, consider the three-state system $n=3$, and let $\alpha_1=\alpha_2^*=e^{i\phi}$. There are three possible values of $s^{*}_j s_k$, given by $1,\omega, \omega^2=\overline{\omega}$.  When $\phi=0$ so that the $\alpha$ are real, the three values of $s^{*}_j s_k e^{i\phi}$
are illustrated in the left part of fig.\ \ref{fig:chiralrotate}.
\begin{figure}[ht]
 \begin{center}
 \includegraphics[scale=0.6]{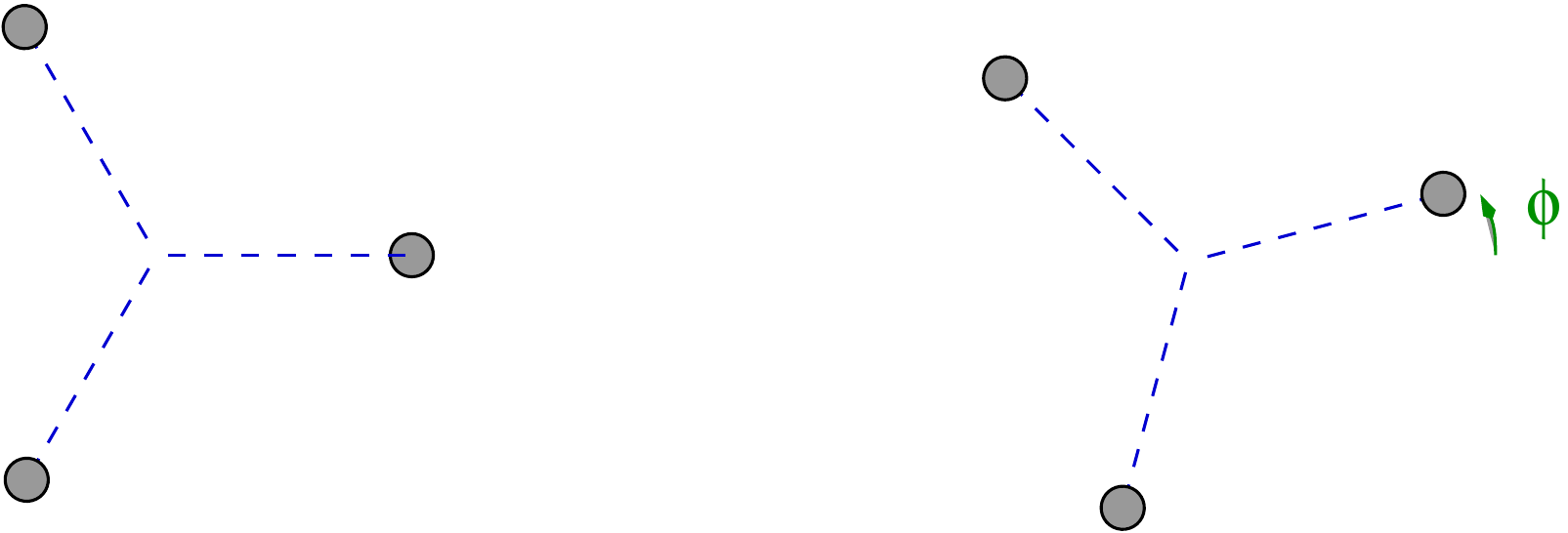}
\end{center}
 \caption{Illustration of the three possible values of $s^*_j s_k e^{i\phi}$ in the $\zth$ model for $\phi=0$ and $\phi\ne 0$.}
\label{fig:chiralrotate}
\end{figure}
For a given pair of spins, the energy is proportional to minus the real part of each value, so the rightmost point in this figure $s_k=s_j$ minimizes the energy. This indeed is a ferromagnetic interaction when $\phi=0$. Now take $\phi>0$ and small, as illustrated in the right part of fig.\ \ref{fig:chiralrotate}. The interaction still favors alignment, since the rightmost point still has the largest real part. However, a non-zero $\phi$ means that the interaction no longer is invariant if $s_k$ and $s_j$ are interchanged, e.g.\ the spins $(s_j,s_k)=(1,\omega)$ do not give the same energy as $(\omega,1)$ do. Indeed, in the figure the two points on the left do not have the same real part. This means that for $\phi\ne \pi/3$ times an integer, spatial parity symmetry in any direction is broken. For this reason, a model allowing complex $\alpha$ is referred to as a {\em chiral clock} model \cite{Ostlund,Huse}.
 
In the chiral clock model it therefore is possible to continuously interpolate between ferromagnet and antiferromagnet without changing $f$ or $J$. The antiferromagnetic case corresponds to $\phi=\pi/3$ (times any non-zero integer), where there are (at least) two values of $s^*_j s_k$ that minimize the energy, as illustrated in the rightmost part of fig.\ \ref{fig:chiralrotate2}.
\begin{figure}[ht]
 \begin{center}
 \includegraphics[scale=0.6]{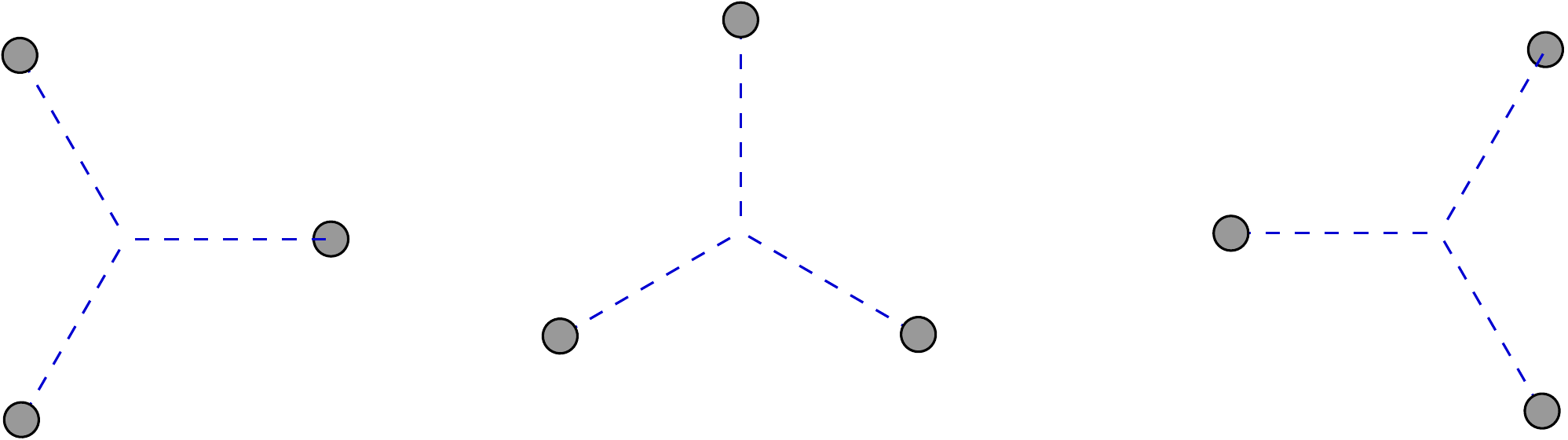}
\end{center}
 \caption{Illustration of the three possible values of $s^*_j s_k e^{i\phi}$ for the $\zth$ ferromagnetic case $\phi=0$, the ``symmetric'' case $\phi=-\pi/6$ and the antiferromagnetic case $\phi= \pm\pi/3$.}
\label{fig:chiralrotate2}
\end{figure}
A particularly interesting case is $\phi=\pm\pi/6$ (or $\pi/2$),  ``halfway'' between ferromagnetic and antiferromagnetic. This is illustrated in the middle picture in fig.\ \ref{fig:chiralrotate2}. As is apparent, this case is {\em symmetric} around the imaginary axis, so that for each state of energy $E$, there is one of energy $-E$. With the quantum Hamiltonian $H$ to be discussed shortly, this means its spectrum is invariant under sending $H\to -H$. The symmetric case of the chiral clock model shares this property with the Ising model, as discussed in the previous section.

\subsection{The Hilbert space and the Hamiltonian}

Taking anisotropic couplings in the two-dimensional chiral clock model, one can define a quantum clock Hamltonian acting on a chain of $L$ ``spins'' with $n$ states each.  The Hilbert space for these $\zn$-invariant chains is thus 
$(\mathbb{C}^n)^{\otimes L}$. The basic operators $\sigma$ and $\tau$  generalize the Pauli matrices $\sigma^z$ and $\sigma^x$ to an $n$-dimensional space. Instead of anticommutation relations, the operators satisfy
\begin{eqnarray}
\label{znspin1}
\sigma^n = \tau^n =1\ , \qquad \sigma^\dagger &=& \sigma^{n-1}\ , \qquad \tau^\dagger = \tau^{n-1}\ ,\\
\sigma\tau &=& \omega\, \tau\sigma\ ,
\label{znspin2}
\end{eqnarray}
where $\omega\equiv e^{2\pi i/n}$.  
Although an explicit representation of these operators is not necessary for the subsequent analysis, it can be useful to keep in mind the following one. Diagonalizing one of these two operators  (say $\sigma$) gives
\be \sigma = 
\begin{pmatrix}
1&0&0&\ \dots\  & 0\\
0&\omega&0&\ \dots\  & 0\\
0&0&\omega^2&\   & 0\\
\vdots&\vdots&&&\vdots \\
0&0&0&\ \dots\  & \omega^{n-1}
\end{pmatrix},\quad\quad
 \tau = 
\begin{pmatrix}
0&0&0\ \dots\  &0& 1\\
1&0&0\ \dots\  &0 & 0\\
0&1&0\ \dots   & 0 &0\\
\vdots&&&\vdots&\vdots \\
0&0&0\ \dots\ & 1& 0
\end{pmatrix}
\label{explicitrep}
\ee
In this representation $\sigma$ measures the value of the spin at each site, while $\tau$ shifts the spin. The Hamiltonian is defined in terms of operators $\sigma_j$ and $\tau_j$ acting non-trivially at site $j$ of the chain. Each pair of operators $(\sigma_j,\,\tau_j)$ satisfies the algebra (\ref{znspin1},\ref{znspin2}), while operators at different sites commute.

The natural generalization of the Ising Hamiltonian to the three-state case for open boundary conditions is
\be
H_{3} = -f \sum_{j=1}^L \left(\tau^\dagger_j e^{-i\phi} + \tau_j e^{i\phi}\right) 
\ -\ J \sum_{j=1}^{L-1} \left(\sigma_j^\dagger\sigma_{j+1} e^{-i\phih} + \sigma_j\sigma_{j+1}^\dagger e^{i\phih}\right) 
 \label{H3}
\ee
There are thus three physically important parameters in $H_3$: $f/J$, $\phi$ and $\phih$. 
The one-site term, with the $f$ in front, generalizes the flip term, whereas the two-site term with $J$ in front generalizes the nearest-neighbor interaction. When the phases $\phi=\phih=0$, this is exactly the quantum chain version of the three-state Potts model. The point $f=J$, $\phi=\phih=0$ is self-dual, critical, and integrable \cite{TL}. It separates the ordered phase, where the $\zth$ symmetry is spontaneously broken, from a disordered phase. The spectrum is invariant under $H_3\to -H_3$ in the symmetric case $\phi=\phih=\pi/6$ (plus $\pi/3$ times any integer). 
%
 

For non-vanishing phases, the physics of the model is much more intricate (see e.g.\ \cite{Ostlund,Huse,Perkoverview}). Although this is quite interesting, only a few simple properties will be necessary for the subsequent analysis.
The discussion in the previous subsection \ref{sec:classicalPotts}  of behavior of the classical model when the phase is changed applies here as well. Varying the phases $\phi$ and $\phih$  allows one to interpolate between the ferromagnet and antiferromagnet at constant $J$. Here, as opposed to the Ising chain, the antiferromagnetic and ferromagnetic cases are {\em not} equivalent, as apparent in fig.\  \ref{fig:chiralrotate2}; there is only one way for nearest-neighbor spins to align ferromagnetically, but there are (at least) two for the antiferrogmagnet. Outside of the ferromagnetic and antiferromagnetic cases, the chiral Hamiltonian breaks spatial parity symmetry and time-reversal symmetry.

For general $n$, the chiral clock Hamiltonian with open boundary conditions is
\be
H_n = -f \sum_{j=1}^L \sum_{m=1}^{n-1} \alpha_m (\tau_j)^m 
\ -\  J\sum_{j=1}^{L-1} \sum_{m=1}^{n-1} \ah_m (\sigma_j^\dagger \sigma_{j+1})^m\ . 
\label{Hn}
\ee
With $f$ and $J$ real and non-negative by definition, the conditions on the couplings
\begin{equation}
\alpha^*_m= \alpha_{n-m}\ , \qquad \ah^*_m = \ah_{n-m} \ 
\label{Hherm}
\ee
are necessary to make the Hamiltonian hermitian.
The Hamiltonian is invariant under the $\zn$ symmetry of sending $\sigma_j \to \omega \sigma_j$ for all $j$, i.e.\ ``increasing'' each spin. The corresponding  symmetry generator, generalizing $(-1)^F$,  is
\be
\omega^P\equiv \prod_{j=1}^L \tau_j^\dagger\ ,
\ee
which indeed satisfies $(\omega^P)^n_{} =1$.

\subsection{The chiral clock model in terms of parafermions}

Parafermions in an $n$-state spin chain are defined analogously to the fermions in the Ising/Majorana chain, by multiplying order and disorder parameters \cite{FK}. At each site there are two basic parafermions $\chi_j$ and $\psi_j$, generalizing $a_j$ and $b_j$:
\begin{eqnarray}
\chi_j &=& \left(\prod_{k=1}^{j-1} \tau_k\right) \sigma_j \ ,\\
\label{chidef}
\psi_j &=&\omega^{(n-1)/2}\,\chi_j\tau_j= \omega^{(n-1)/2}\, \left(\prod_{k=1}^{j-1}\tau_k\right)  \sigma_j\tau_j\ .
\label{psidef}
\end{eqnarray}
Like $\sigma$ and $\tau$, these do not square to 1 and do not commute, but rather
\begin{eqnarray}
(\chi_j)^n = (\tau_j)^n =1\ , \qquad \chi_j^\dagger &=& (\chi_j)^{n-1}\ , \qquad \psi_j^\dagger = (\psi_j)^{n-1}\ ,\\
\chi_j\psi_j &=& \omega \,\psi_j\chi_j\ .
\label{chipsij}
 \end{eqnarray}
Because of the strings attached, operators at different points do not commute like $\sigma_j$ and $\tau_k$ do. Instead, 
\begin{equation}
\chi_{j}\chi_k = \omega\, \chi_k \chi_j\ ,\qquad \psi_j\psi_k = \omega\, \psi_k\psi_j\ ,\qquad \chi_j\psi_k = \omega\, \psi_k\chi_j\qquad 
\hbox{ for } j<k.
\label{chipsijk}
\ee
The restriction $j<k$ is necessary for these relations to make sense; only for Ising is $\omega=\omega^{-1}$.  Thus while parafermions are a natural generalization of fermions, calculations involving them are much more intricate.

The chiral clock Hamiltonian (\ref{Hn}) can be written simply in terms of parafermions. By definition,
\[ \tau_j=\omega^{-(n-1)/2} \chi^\dagger_j\psi_j\ , \qquad \sigma^\dagger_j\sigma_{j+1} = 
\omega^{-(n-1)/2}  \psi_j^\dagger\chi_{j+1}\ .\]
Using these along with the parafermionic commutation relation (\ref{chipsij}) means that 
\be
H_n = -f \sum_{j=1}^L \sum_{m=1}^{n-1} \alpha_m\, \omega^{m(m-n)/2}\, (\chi_j)^{n-m}(\psi_j)^m
-  J\sum_{j=1}^{L-1} \sum_{m=1}^{n-1} \ah_m \,\omega^{m(m-n)/2} \, (\psi_j)^{n-m}(\chi_{j+1})^m\ . 
\label{Hnp}
\ee
The extra factors of $\omega$ look strange, but they ensure that $H_n$ is Hermitian; note e.g.\ that $(\chi_j^\dagger \psi_j)^\dagger = \psi_j^\dagger\chi_j = \omega \chi_j \psi^\dagger_j$.

The three-state case is
\be
H_{3} = -f \sum_{j=1}^L \left(\chi^\dagger_j \psi_j \, \alpha_1\overline{\omega} + \hbox{ h.c. }\right) 
\ -\ J \sum_{j=1}^{L-1} \left(\psi_j^\dagger\chi_{j+1}\, \ah_1\overline{\omega} + \hbox{ h.c }\right) \ .
 \label{H3f}
\ee
The symmetric case $\alpha_1 = \ah_1 = e^{-i\pi/6}$ has a particularly simple form: 
\[
H_{3} = if \sum_{j=1}^L \left(\chi^\dagger_j \psi_j\ -\ \psi_j^\dagger\chi_{j}\right)
\ +\ iJ \sum_{j=1}^{L-1} \left(\psi_j^\dagger\chi_{j+1}\ -\ \chi^\dagger_{j+1} \psi_j\right) \ .
\]

Even though $H_n$ looks simple in terms of the parafermions, it does not allow for any easy computation of its spectrum as with the Ising chain. The fundamental reason why is in the relations (\ref{chipsijk}). One can of course Fourier transform the parafermions into momentum space as with the fermionic solution of the Ising chain. The catch here is that because of the $j<k$ requirement in (\ref{chipsijk}),  commutation relations of parafermions at fixed momentum couple different momenta. Thus Fourier transforming  the Hamiltonian does not reduce it into $4\times 4$ blocks, the way it does for Ising.

The chiral clock chain is however integrable for a two-parameter family of couplings \cite{Gehlen,AMPTY}. This integrable case is referred to as the ``integrable chiral Potts chain'' (even though there is only $\zn$ symmetry, not $S_n$); overviews can be found in \cite{Perkoverview,Baxteroverview}. The two independent parameters $\phi$ and $\phih$ are defined by
\be
\alpha_m = \frac{1}{\sin(\pi m/n)} e^{i\phi(2m-n)} \ , \qquad \ah_m = \frac{1}{\sin(\pi m/n)} e^{i\phih(2m-n)}\ ,
\label{alphaphi}
\ee
while the flip and interaction coefficients are
\be
f\cos(n\phi) = J\cos(n\phih)\ .
\label{fJphi}
\ee
The ``superintegrable'' case $\phi=\phih=\pi/(2n)$, where any value of $f/J$ is allowed, is particularly interesting \cite{Howes,Gehlen}. In particular, it resembles the Ising/Majorana chain in that the terms in the Hamiltonian obey the Onsager algebra \cite{Onsager1}, the key technical observation that allowed Onsager to solve the Ising model without utilizing the Jordan-Wigner transformation later exploited by Kaufmann. Another interesting resemblance in that its spectrum is invariant under sending $H\to -H$.  It turns out that the parafermions have a nice relation to integrability; for example the relations (\ref{alphaphi},\ref{fJphi}) giving integrability also arise by demanding that there exist a parafermionic raising or lowering operator, or ``shift mode''. I discuss this in a companion paper \cite{me}.

\section{Parafermionic edge zero modes}
\label{sec:edge}

In this section I discuss an iterative procedure for finding the edge zero mode in these $\zn$-symmetric systems. This procedure works only when the interactions are chiral, i.e.\ the chain is {\em not} a simple ferromagnet or antiferromagnet.   A striking result, therefore, is that only in when spatial-parity and time-reversal symmetries are broken does an exact edge mode exist by this construction. In the next section, I prove that for the appropriate couplings this procedure works to all orders, thus showing that this edge zero mode is exact.

A {\em parafermionic zero mode} $\Psi$ is an {\em operator} that commutes with the Hamiltonian but not with the $\zn$ charge $\omega^P$:
\be 
[H_n,\Psi]=0,\qquad \omega^P \Psi = \omega\, \Psi\omega^P\ . 
\ee
To survive in the $L\to\infty$ limit, it needs to be normalizable:  $\Psi^\dagger\Psi =1$.
The fact that it does not commute with $\omega^P$ means that it maps between different $\zn$ sectors.
If such a mode exists, the spectrum therefore must be the same in all sectors. This therefore is a much stronger statement than simply whether a zero-energy eigenstate exists; it is a statement about the entire spectrum.

To illustrate this, the energy spectrum in the $\zth$ symmetric case ($\phi=\phih=\pi/6$ in (\ref{H3f})) with $f=J/2$ is plotted in fig.\ \ref{fig:energylevels} for open boundary conditions and just $L=4$ sites. Quite obviously, the spectrum in each of the three $\zth$ symmetry sectors is very close, consistent with the existence of a zero mode. As $L$ is increased, the match between the energies becomes exponentially closer.
\begin{figure}[ht]
 \begin{center}
 \includegraphics[scale=0.6]{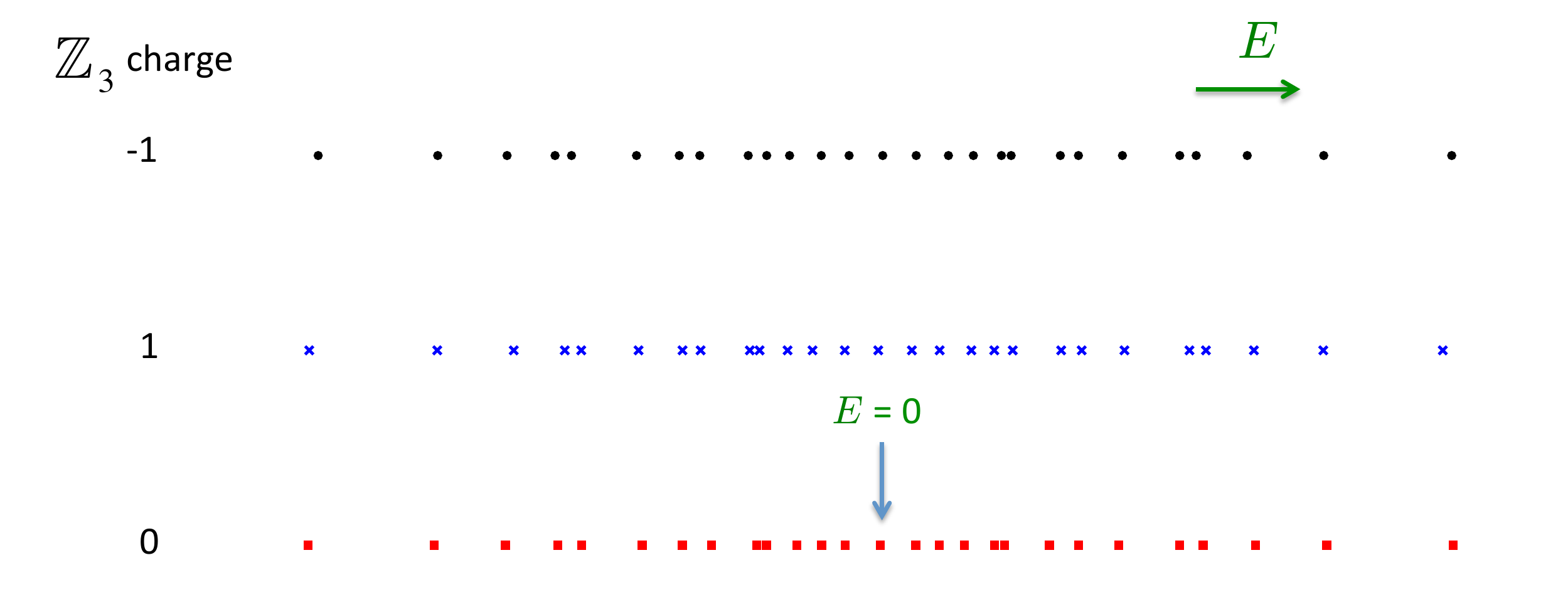}
\end{center}
\vskip-.2in
 \caption{The energy levels in the symmetric $\zth$ case with $f=J/2$ and $L=4$ and open boundary conditions, plotted so increasing energy is to the right. The overall energy scale is unimportant; the purpose of this plot is to illustrate the close correspondence between the spectra in different $\zth$ sectors.}
\label{fig:energylevels}
\end{figure}
The alignment starts out perfectly at $f/J=0$ where the edge zero mode is exact even at finite size, and remains excellent as $f/J$ is increased until around $.5$. For higher values of $f/J$, the agreement noticeably worsens. It is hard to say whether a transition has occurred, or this is simply a finite-size effect. Going to large sizes means that there are far more levels, and so to distinguish them one must ``zoom in'' on a much small region of energies, thus making it difficult to distinguish whether these are becoming closer or not in the continuum limit. However, this plot strongly suggests that a zero mode survives in the symmetric case at least until $f/J\approx 1/2$, and very probably further.

\vskip.2in
\begin{figure}[ht]
 \begin{center}
 \includegraphics[scale=0.6]{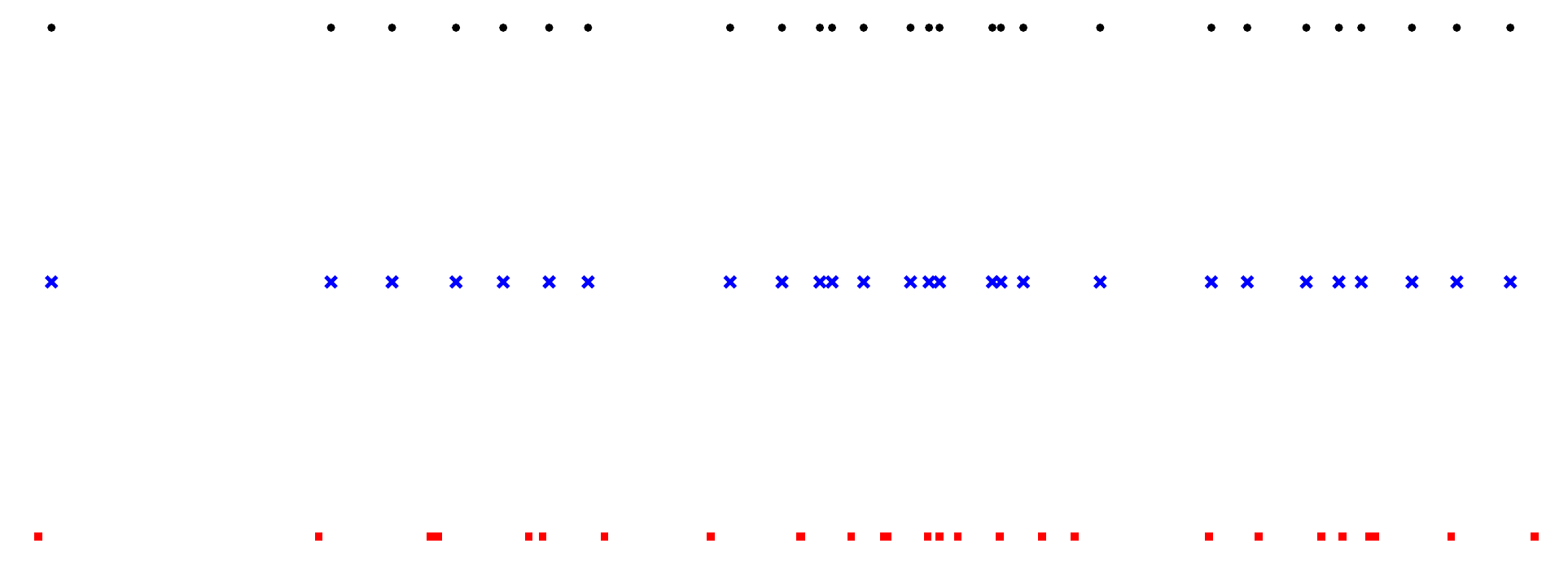}
\end{center}
\vskip-.2in
 \caption{The energy levels in the $\zth$ ferromagnetic case with $f=J/2$ and $L=4$, plotted in the same fashion as in figure \ref{fig:energylevels}. The energy levels for the antiferromagnetic case are given by sending $E\to -E$.}
 \label{fig:energylevels0}
\end{figure}
One the other hand, in the ferromagnetic $\zth$ case $\alpha=\ah=0$ plotted in figure \ref{fig:energylevels0} for $f=J/2$, the correspondence between levels in different sectors is much less strong. (The fact that the spectrum is identical for $\zth$ charges $1$ and $-1$ has nothing to do with a zero mode, but instead is a consequence of the $S_3$ permutation symmetry present only  at the ferromagnetic and antiferromagnetic points.) Varying $f/J$ shows that the close correspondence seems to break down for very small $f/J$.  Although some of the levels in the figure are very close in value, others differ by more than 10\% between sectors. This is a strong indication that there is no exact zero mode in the ferromagnetic case except perhaps at very small values of $f/J$. This conclusion also applies to the antiferromagnetic case $\phi=\phih=\pi/3$, since the spectrum is that of the ferromagnetic case with a sign flip.

The existence of an exact edge zero mode is easy to show in the extreme case $f=0$. In this limit, the spins are ordered (except at the antiferromagnetic points). Namely, each interaction energy
\[ -J \sum_{m=1}^{n-1} \alpha_m (\sigma_j^\dagger\sigma_{j+1})^m\]
commutes with those at other values of $j$, and so all their eigenvalues can be minimized. Unless the $\alpha_m$ are chosen to give multiple minima (the antiferromagnet), fixing the spin at site $1$ then fixes all the other spins in the ground state.  There are then $n$ exactly degenerate ordered ground states, with each of the possible eigenvalues $1,\omega,\dots \omega^{n-1}$ of the $\zn$ symmetry generator $\omega^P$. 
In fact, at $f=0$, the entire spectrum is the same in any of the $n$ sectors. The parafermions give a simple way of showing this. Similarly to the fermionic case, the operators  $\chi_1$ and $\psi_L$ do not appear in the Hamiltonian with open boundary conditions. Since all the terms in the Hamiltonian are of the form $\chi_j^{n-m}\psi_k^m$ with $1<j\le L$ and $1\ge k < L$,  the parafermionic commutation relations (\ref{chipsij},\ref{chipsijk}) ensure that each commutes with both $\chi_1$ and $\psi_L$.  Thus
\[ [H_n (f=0),\chi_1] =  [H_{n} (f=0),\psi_L] = 0\ .\]
Since 
\[\omega^P \chi_1 = \omega\, \chi_1 \omega^P\ , \qquad \omega^P \psi_L = \omega\, \psi_L \omega^P \]
these operators change $\zn$ sectors. $\chi_1$ and $\psi_L$ are obviously localized at the edges, and each taken to the $n$th power gives $1$. Thus $\chi_1$ and $\psi_L$ are indeed exact edge zero modes.

Since there is a gap, one expects that the zero mode will remain at least approximately a zero mode for small enough $f/J$. However, that is not a convincing demonstration of topological order. One would like to see that it remains a zero mode, exact in the $L\to\infty$ limit, in at least some range of $f/J$. This requires showing that is normalizable; recall that in the Ising/Majorana case described in section \ref{sec:edgeIsing}, the would-be edge zero mode is no longer normalizable for $f\ge J$. This is a signal that the phase transition is taking place. The phase diagram of the chiral clock model however is much more complicated \cite{Ostlund,Huse,Howes,AMP,Perkoverview}, and so it is not at all obvious such a simple picture holds. In general, without an explicit expression a precise range of $f/J$ where the zero mode survives in the $L\to\infty$ limit.  However, in this section I show that the more breaking of parity, the farther it survives. The symmetric case, ``halfway'' in between ferromagnet and antiferromagnet, is the case where the edge zero mode seems most robust.

The major complication in demonstrating this explicitly is the fact that commuting $H_n$ with a parafermion does not automatically preserve ``parafermion number'' like it does for fermions. Namely, note from (\ref{Hacomm},\ref{Hbcomm}) that commuting the Ising Hamltonian with something linear in the $a_j$ and $b_j$ gives something linear in the $a_j$ and $b_j$. However, in the more general parafermion case, one has e.g.\
\[ [\chi_j^\dagger\psi_j, \psi_j]\  \propto\ \chi_j^\dagger\psi^\dagger_j\ . \]
Put more prosaically, this means that there is no such thing as a free parafermion.

The obvious way to try to construct the edge zero mode is iteratively, as with the Ising/Majorana chain. Namely, break the Hamiltonian with open boundary conditions  (\ref{Hn}) or (\ref{Hnp}) into two pieces $H_n=F + V$, where $F$ contains all the flip terms (those with an $f$ in front), and $V$ all the interaction terms (those with a $J$). The edge parafermion $\chi_1$ commutes with $V$, which is why it is a zero mode in the $f=0$ case. It does not commute with $F$, but instead gives
\be [H_n,\chi_1]= [F,\chi_1] = f\sum_{m=1}^{n-1} \alpha_m \omega^{m(m-n)/2}(1-\omega^{-m})(\chi_1)^{n-m+1}(\psi_1)^m \ .
\label{firstit}
\ee
The question is then: can this be written as the commutator of $V$ with something else? If the answer is yes, then this something else can be subtracted from $\Psi$, so that the commutator $[H_n,\Psi]$ yields only terms of order $f^2$. One can then commute with $F$ again, and attempt again to write the result as a commutator of $V$. If this iteration can be successfully repeated to all orders, then $\Psi$ is indeed an exact edge zero mode.

For simplicity consider first the $\zth$ case, where
\[ [F,\chi_1] = 2if\left(e^{-i\phi} \psi_1 - \omega e^{i\phi} \chi_1^\dagger\psi_1^\dagger\right)\ ,\]
with $\alpha_1=\alpha_2^* = e^{-i\phi}/\sin(\pi/3)$. 
Since $V$ does not involve $\chi_1$, for the iteration procedure to work each of these two terms individually must result from a commutator with $V$: 
\[ [V, X] = \psi_1 \ ,\qquad [V,Y] = \chi_1^\dagger\psi_1^\dagger\ .\]
The only terms in $V$ possibly giving $\psi_1$ on the right-hand side are those  involving $\psi_1^\dagger\chi_2$ and  $\psi_1\chi_2^\dagger$.
Thus $X$ must be some linear combination
\[ X = A\, \psi_1 + B\,\chi_2 + C\,\psi_1^\dagger\chi_2^\dagger\ ,\]
because commuting with $V$ maps such a linear combination onto another such linear combination: $[V,X]= A' \psi_1 + B'\chi_2 + C'\psi_2^\dagger\chi_2^\dagger$. This map from $(A,B,C)$ to $(A',B',C')$ is linear, and so it can be written in matrix form as
\be
\begin{pmatrix}
A'\\ B'\\ C'\\
\end{pmatrix}
= 
2iJ
\begin{pmatrix}
0 & -e^{i\phih} & e^{-i\phih}\overline{\omega}\\
e^{-i\phih}& 0 &-e^{i\phih}\overline{\omega}\\
-e^{i\phih}\omega& e^{-i\phih}\omega & 0
\end{pmatrix}
\begin{pmatrix}
A\\ B\\ C\\
\end{pmatrix}
\label{matrixcomm}
\ee
using the parametrization $\ah_1=\ah_2^* = e^{-i\phih}/\sin(\pi/3)$. Answering the question is now simple: there is an $X$ satisfying $[V, X]=\psi_1$ if this matrix can be inverted. The determinant is $-16J^3\sin(3\phih)$, which is non-zero unless $\phih\ne \pi/3$ times an integer. Thus it is invertible and $X$ can be found for all $\phih$ except for the ferromagnetic and antiferromagnetic cases! 

The analogus matrix giving $Y$ is minus the Hermitian conjugate, and inverting them gives
\[ X= \frac{1}{4J\sin(3\phih)} (\psi_1 + e^{2i\phih}\chi_2 + e^{-2i\phih} \omega\, \psi^\dagger_1\chi_2^\dagger)\ ,\quad
Y= -\frac{1}{4J\sin(3\phih)} (\psi_1^\dagger + e^{-2i\phih}\chi_2^\dagger + e^{2i\phih} {\omega}\, \psi^\dagger_1\chi_2^\dagger)\ .
\]
Therefore the edge zero mode is to order $f/J$:
\be
\Psi_{\rm left} = \chi_1 - 2if\, e^{-i\phi} X + 2if\,e^{i\phi}\chi_1^\dagger Y + \dots\ .
\label{Psifirst3}
\ee
The dimensionless expansion parameter here seems not to simply be $f/J$ as in the fermion case, but rather 
\[ \frac{f}{2J \sin(3\phih)}\ .\] The radius of convergence of this expansion is therefore $\phih$-dependent, and its maximum value is at the symmetric points $\phih=\pi/6$ (plus $\pi/3$ times any integer). Thus the symmetric coupling seems to be the most robust for the presence of topological order.

The fact that the iterative procedure does not work at the parity-invariant ferromagnetic and antiferromagnetic points are approached is not a fluke of the $n=3$ case; it is true for all $n$. The most convenient method of demonstrating this is to show that the matrices generalizing that in (\ref{matrixcomm}) have zero eigenvalues. Since commuting with $H_n$ preserves $\zn$ charge, there are $n$ such matrices, i.e.\
\be
[V, \psi_j^{p}\chi_{j+1}^{Q-p}] =  J \sum_{p'=1}^n \JQ_{pp'} \, \psi_j^{p'}\chi_{j+1}^{Q-p'}\ 
\label{Jdef}
\ee
for $Q=0,1,\dots n-1\ .$
Each of these matrices $\JQ$ is $n\times n$. The commutators necessary follow from 
\be
[\psi_j^{n-m}\chi_{j+1}^m,\psi_j^{p}\chi_{j+1}^q] = 
(\omega^{-mp} -\omega^{mq}) \psi_j^{n-m+p}\chi_{j+1}^{m+q} \ .
 \label{psipsicomm}
 \ee
The definition of the Hamiltonian in terms of parafermions (\ref{Hn}) and the definition of $\JQ$ (\ref{Jdef}) therefore gives the entries of these matrices to be
\be \JQ_{pp'} =   \omega^{m(m-n)/2}\ah_{m}\,\omega^{-mp}(1 -\omega^{mQ}) \ ,
\label{JQdef}
\ee
where $m=(p-p')$ mod$\,n$. Note that the entire matrix ${\cal J}^{(0)}=0$, because $(\psi_j^\dagger \chi_{j+1})$ and all its powers commute with $V$.

Finding the eigenvalues of the $\JQ$ is straightforward because their eigenvectors are independent of the $\ah_m$ (and are the same for all $Q$). Labeling these eigenvectors by $v^{(r)}$ with $r=0,1\dots n-1$, they have elements
\be v^{(r)}_q = \omega^{qr}\omega^{q(n-q)/2} ,
\label{vQr}
\ee
so that the eigenvalue equation $\sum_{p'} \JQ_{pp'} v^{(r)}_{p'} = \lambda^{(Q,r)}
v_p^{(r)}$ yields
\be
\lambda^{(Q,r)} = \sum_{m=1}^{n-1} \ah_m\, \omega^{-mr}(1-\omega^{mQ})\ .
\label{lQr}
\ee
For the three-state case, this gives for $\ah_1=\ah_2^*=e^{-i\phih}/\sin(\pi/3)$,
\[\lambda^{(1,r)} =  4\sin\left(\phih+\frac{2\pi}{3}(r+1)\right)\ \qquad \hbox{ for }n=3 ,\]
as can be checked directly using (\ref{matrixcomm}). 

Terms involving $\psi_1^Q$ (with no $\chi_2$) appear for all $Q\ne 0$ on the right-hand-side of (\ref{firstit}).
Thus if any one of the $\lambda^{(Q,r)} =0$ for $Q\ne 0$, then the iterative procedure will not work.  For the three-state case, this is zero only in the ferromagnetic and antiferromagnetic cases $\phih=\pi/3$ times an integer, where parity is conserved. For higher $n$, one of the $\ah_m$ can always be chosen to make one of the eigenvalues vanish. For example, for $n=4$, one of $\lambda^{(1,r)}=0$ if $\ah_2 = \pm {\rm Re}(\ah_1) \pm {\rm Im}(\ah_1)$.

In the ferromagnetic or antiferromagnetic cases, where all $\alpha_m$ can be taken to be real, at least one of the eigenvalues vanishes. This is easy to see from (\ref{lQr}).
Because $\ah_m = \ah^*_m = \ah_{n-m}$ here, the coefficient of $J\ah_m$ for any $m\ne n/2$ is
\[ \omega^{-mr}(1-\omega^{mQ}) + \omega^{mr} (1-\omega^{-mQ})\ . \]
Any time $Q=2r$, this vanishes for all $m$. Moreover, the coefficient of $\ah_{n/2}$ is proportional to $1-\omega^{nQ/2}=1-(-1)^Q$, so this term vanishes for all even $Q$ as well. Thus
\[ \lambda^{(2r,r)} = 0\  \qquad \hbox{ for }\ah_m \hbox{ real }.\]
If parity and time-reversal symmetry are conserved, the iterative procedure for finding the zero mode does not work! This is reminiscent of the two-dimensional non-Abelian states found in \cite{BarkWen}, where it is necessary to violate time-reversal symmetry to have a $\zn$ topological phase.

\section{The proof that the edge zero modes are exact}
\label{sec:proof}

In the preceding section I discussed edge zero modes, starting an iterative procedure for computing them explicitly. In this section, I show that this procedure works to all orders, provided that the matrices $\JQ$ introduced in the previous section are invertible for $Q\ne 0$. This implies that there exists an edge zero mode for sufficiently small $f/J$. I obtain a closed-form expression for only the ``leading'' piece of the zero mode, the terms that at order $(f/J)^{l-1}$ involve $\chi_l$. Nevertheless, this is enough information to prove that an edge zero mode exists. An added feature is that this proof applies when the coupling vary in space, i.e.\ the Hamiltonian is
\be
H_n = - \sum_{j=1}^L \sum_{m=1}^{n-1} f_j \,\alpha_m (\tau_j)^m 
\ -\  \sum_{j=1}^{L-1} \sum_{m=1}^{n-1} J_j\, \ah_m (\sigma_j^\dagger \sigma_{j+1})^m\ . 
\label{Hnagain}
\ee
This indicates that the edge zero mode, and presumably the topological order, is robust in the presence of disordered couplings.

It is convenient to convert the problem of finding a zero mode of $H_n$ to finding a zero eigenvalue of an associated ``Hamiltonian'' ${\cal H}$.  Because the zeroth order term in $\Psi$ is a parafermion and the Hamiltonian can be expressed in terms of parafermions, the entire series can be expressed in terms of parafermions. 
Thus the edge zero mode can be written in terms of the parafermion operators, i.e.\ is a vector in
the vector space ${\cal V}$ spanned by all the parafermions and their products.  Since $\chi_j^n=\psi_j^n=1$ for all sites $j$, ${\cal V}$ is of dimension $n^{2L}$. Ordering the parafermions as $\chi_1,\psi_1,\chi_2,\psi_2,\dots$, each basis element can be labeled as 
\[
 |p_1 p_2 p_3\dots p_{2L}\rangle\rangle \ \equiv\  \chi_1^{p_1} \psi_1^{p_2} \chi_2^{p_3}\psi_2^{p_4}\dots \psi_L^{p_{2L}} \ .
 \]
 where $\mu = 0\dots n^{2L}$--$1$ with all labels $p_i$ are interpreted mod $n$. 
The commutator of the Hamiltonian with any combination of parafermions can be converted into a {\em linear} operation on ${\cal V}$, as was done in the last section with (\ref{matrixcomm}) and (\ref{Jdef}). 
Namely, any commutator of $H_n$ with any element $v_{\mu}$ of ${\cal V}$ (i.e.\ any linear combination of any products of parafermions)  with the Hamiltonian can be written as
\be
[H_n, v_{\mu}] = {\cal H}_{\mu\nu} v_{\nu} 
\label{Hv}
\ee
for a $n^{2L}\times n^{2L}$ matrix ${\cal H}$. An operator comprised of parafermions commuting with $H$ (i.e.\ a zero mode) therefore corresponds to an eigenvector of ${\cal H}$ with zero eigenvalue.

The explicit form of the matrix ${\cal H}$ can be found by using (\ref{Hacomm},\ref{Hbcomm}) in the fermionic case, while the general case requires (\ref{psipsicomm}) and
\[
 [\chi_j^{n-m}\psi_j^m,\chi_j^{p}\psi_j^q] = (\omega^{-mp} -\omega^{qm}) \chi_j^{n-m+p}\psi_j^{m+q}\  \ .
 \]
The matrix ${\cal H}$ then can be written as
\be
{\cal H} = -{\cal F}  - {\cal J}\ ,
\ee
\[ {\cal F}= \sum_{j=1}^{L} f_j\, {\cal H}_{2j-1}\ , \qquad {\cal J} =\sum_{j=1}^{L-1} J_j\, {\cal H}_{2j}\ .
\]
where ${\cal H}_k$ acts non-trivially on the ``state''
\[v_{\mu} = |\dots p,q\dots\rangle\rangle\ \]
with $p$ and $q$ in the $k$th and $k+1$st places respectively, taking it to the states
\[v_{\nu}= |\dots p-m, q+m\dots\rangle\rangle\]
with $m=1,\dots n-1$.
The corresponding non-zero matrix elements for $k$ even are those from (\ref{JQdef}):
\be ({\cal H}_{2j})_{\mu\nu} =   \omega^{m(m-n)/2}\ah_{m}\,(\omega^{-mp} -\omega^{mq}) \ ,
\label{calHdef}
\ee
while those for $k$ odd are given by replacing $\ah_m$ with $\alpha_m$:
\be ({\cal H}_{2j-1})_{\mu\nu} =   \omega^{m(m-n)/2}\alpha_{m}\,(\omega^{-mp}-\omega^{mq}) \ .
\label{calHdef2}
\ee
 Because the $m=0$ term does not appear above, there are no diagonal elements in ${\cal H}$; ${\cal H}_{\mu\mu}=0$, as is easily checked in (\ref{calHdef},\ref{calHdef2}). 
 
The matrix ${\cal H}_{\mu\nu}$ is hermitian:
using (\ref{calHdef2}) with $\mu$ and $\nu$ reversed gives
\[{\cal H}_{\nu\mu} =  \omega^{m(m+n)/2}\alpha_{n-m} \omega^{m(p-m)}(1 -\omega^{m(q-p)})\ =\ {\cal H}_{\mu\nu}^*.
\]
because $\alpha_{n-m}=\alpha_m^*$ and $\omega^{\pm mn/2}=(-1)^m$. Moreover, ${\cal V}$ can be made into a Hilbert space by defining the inner product
\[ \langle\langle p_1 p_2\dots p_{2L}| q_1 q_2\dots q_{2L} \rangle\rangle= \delta_{p_1q_1} \delta_{p_2q_q}\dots \delta_{p_{2L}q_{2L}}\ .\]
Note that the adjoint of $v_\mu$ in this new inner product is {\em not} the conjugate of products of parafermions that was used to define this basis; it is just a formal definition that makes ${\cal H}$ self-adjoint.

Thus ${\cal H}$ can be thought of as a ``Hamiltonian'' of a $n^{2L}$-dimensional quantum-mechanical system (recall the original system has dimension $n^L$).  Tools such as perturbation theory familiar from quantum mechanics can be used to study the zero-``energy'' states of the new system, i.e.\ the zero modes of the original system. In the appendix, ${\cal H}$ is written out in more standard form.
This new ``Hamiltonian'' is quite reminiscent of the original Hamiltonian, seeing as it acts on a Hilbert space with $n$ states per site, has a $\zn$ symmetry, and if not for the term in parentheses, the coefficients would be the same. The appendix details how this ``Hamiltonian'' can be split into two commuting copies of the original (with opposite signs in front).

I showed in the previous section that the iteration procedure works to at least to order $f/J$ if the $\ah_m$ are such that $\JQ$ for $Q\ne 0$ is invertible, i.e.\ none of the eigenvalues $\lambda^{(Q,r)}$ in (\ref{lQr}) are zero. For the remainder of this section, assume that this condition holds. This means that ${\cal H}_{2k}$ can be inverted as long as $p,q$, the $2k,2k+1$ entries in the vector on which it is acting, obey $(p+q)$ mod$\,n\ne 0$. Thus let ${\cal G}_{2k}$ be this inverse, defined in general so that
\be
{\cal G}_{2k} {\cal H}_{2k}|\dots pq\dots\rangle\rangle =
{\cal H}_{2k} {\cal G}_{2k}|\dots pq\dots\rangle\rangle =
  \delta_{0,(p+q)\,{\rm mod}\,n}|\dots pq\dots\rangle\rangle \ .
\label{GH}
\ee
Constructing ${\cal G}_{2k}$ explicitly is straightforward, using the eigenvectors and eigenvalues of $\JQ$ found in (\ref{vQr},\ref{lQr}).

Rephrasing the iteration procedure in terms of the action of the ``Hamiltonian'' ${\cal H}$, the first term $\chi_1$ in the expansion is in the new notation
$v_0\equiv |100\dots\rangle\rangle$. The only term acting on this giving a non-vanishing result is ${\cal H}_1$:
\[{\cal H} v_0 =  f_1{\cal H}_1 v_0  = f_1 \sum_{m=1}^{n-1}\alpha_m \omega^{m(m-n)/2}(\omega^{-m}-1)
 |1-m\, m\, 0\,0\,0\dots\rangle\rangle\ .\]
As in the last section, the question is then if this can be written as ${\cal H}_2$ acting on some state. ${\cal H}_2$ acts non-trivially on the second and third slots, so its invertibility depends on the value of $p_2+p_3=Q$, which is equal to $m$ in this state. Since $m\ne 0$ here, and by assumption ${\cal H}_2$ in the other sectors is invertible, a ${\cal G}_2$ can be constructed obeying (\ref{GH}). Therefore to order $f_1/J_1$, the left zero mode $\Psi$ is 
\be
 \Psi = v_0\ - \ \frac{f_1}{J_1}\, {\cal G}_2{\cal H}_1 v_0\ +\ \dots\ ;
\label{Psifirst}
\ee
acting with $f_1{\cal H}_1 + J_1{\cal H}_2$ on this first-order $\Psi$ gives terms of order $(f_1)^2$.
The first-order expression for $\Psi$ is given explicitly in terms of the parafermions for $n=3$ in (\ref{Psifirst3}).

One can obviously continue in this fashion. However, there is a major potential complication. In the first correction to $\Psi$, there are terms involving $\chi_2$. Thus ${\cal H}_3$ does not annihilate it, and so the next iteration
\be
(f_1 {\cal H}_1 + f_2{\cal H}_3)\,  {\cal H}_3{\cal G}_2{\cal H}_1v_0
\label{nextit}
\ee
involves parafermions in the first four slots, i.e.\ those up to $\psi_2$. This means that the {\em combination} $J_1{\cal H}_2 + J_2{\cal H}_4$ must now be invertible when acting on it. This not at all obviously true. In this particular case, by brute force one can check that it is invertible as long as ${\cal H}_2$ and ${\cal H}_4$ individually are (for $Q\ne 0$). However, not only does brute force quickly become unwieldy, but after a few orders, it doesn't even work!

The problem is familiar from degenerate perturbation theory in quantum mechanics. Let $H=H^{(0)} + H^{(1)}$, with the state $|v_0\rangle$ be the unperturbed state, and $|0_a\rangle$ any states degenerate with it under the unperturbed Hamiltonian $H^{(0)}$. Perturbation theory starting from  $|v_0\rangle$ works  naively only if $\langle 0_a|H^{(1)} |v_0\rangle =0$ for all $a$; otherwise the ``energy denominator'' is zero. The problem persists at higher orders. Denoting the state  at $l$th order in perturbation theory to be $|v_l\rangle$, the naive procedure breaks down if the overlap $\langle 0_a| H^{(1)} |v_l\rangle\ne 0$.

In finding an zero eigenvalue state by this iteration procedure, the problem of inverting ${\cal J} = \sum_j J_{j} {\cal H}_{2j}$ is essentially the same one. It has many zero eigenvalues; for example, it annihilates any  state
$|p_1p_2p_3\dots p_L\rangle\rangle $
that has $p_{2j} + p_{2j+1} = 0$ mod$\,n$ for all $j=1,\dots L-1$.
These zero eigenvalues survive even under perturbation; I show in the appendix that there remain at least $n^L$ of them.  (Although having so many zero modes seems like good sign for having one of them be an edge zero mode, it does complicate the analysis!) Denote any state annihilated by ${\cal J}$ as $|0_a\rangle\rangle$, and let $|v_l\rangle\rangle$ be the order $(f/J)^l$ contribution to $\Psi$. Then if 
\[
\langle\langle 0_a | {\cal F} |v_l \rangle\rangle \ne 0\ 
\]
then ${\cal J}$ cannot be inverted on the appropriate subspace, i.e.\ there exists no $v_{l+1}$ obeying 
\be
{\cal J}\,|v_{l+1}\rangle\rangle = {\cal F}\, |v_l \rangle\rangle\ .
\label{iteration}
\ee

This problem can be fixed in a similar fashion as in degenerate perturbation theory, by ``correcting'' the states order by order. The key point is that $|v_{l+1}\rangle\rangle$ is not uniquely fixed by the requirement (\ref{iteration}), but one can add any state $|0_a\rangle\rangle$ to it while still satisfying (\ref{iteration}). Thus one can always add some linear combination of them to $|v_{l}\rangle\rangle$ make
\be
\langle\langle 0_b |\, {\cal F}\, \Big(|v_l \rangle\rangle- \sum_a C_a| 0_a\rangle\rangle\Big) = 0\ 
\label{goodoverlap}
\ee
for all $b$. The proof of this claim is simple. Since ${\cal J}$ is Hermitian, it has a complete set of orthonormal eigenstates spanning the vector space ${\cal V}$, including those with zero eigenvalue $|0_a\rangle\rangle$. 
Since ${\cal F}$ is Hermitian as well, the action of ${\cal F}$ within the set of states $|0_a\rangle\rangle$ can be diagonalized, i.e.\ the states $|0\rangle\rangle_a$ defined so that $\langle\langle 0_b| {\cal F} |0_a\rangle\rangle =  \Lambda_a \delta_{ab}$. Moreover,  ${\cal F}|v_l\rangle\rangle = \sum_a D_a|0_a\rangle\rangle + \dots$, where the states in the $\dots$ have a non-zero eigenvalue of ${\cal J}$ and so are orthogonal to all the $|0_a\rangle\rangle$. Thus letting $C_a= D_a/\Lambda_a$ for the states with $\Lambda_a\ne 0$, and $C_a=0$ for the others, gives (\ref{goodoverlap}). Order by order, $\Psi$ always can be corrected so that it is an exact zero mode.

There is a catch, however: there is no automatic guarantee that the ``corrected'' state still will be an edge state. 
 (If there were a guarantee, then the zero mode in the ferromagnetic case could be ``corrected'' into an edge one.)  In an edge state, the parafermion $\chi_l$ must appear in terms at least order $(f/J)^{l-1}$ in the expansion of $\Psi$. Nonetheless it is not difficult to show that the extra terms as in (\ref{goodoverlap}) do indeed keep $\Psi$ an edge mode, as long as ${\cal J}^{(Q)}$ is invertible for $Q\ne 0$. 

The key observation is that the iteration can be employed without any corrections to get the leading terms at each order. The leading terms in an edge mode at order $(f/J)^{l-1}$ are those involving the parafermion $\chi_l$, i.e.\ contain only states with $p_{2l-1}\ne 0$ and $p_k\ne 0$ for $k > 2l-1$. I refer to such states as being of ``length'' $2l-1$. At zeroth order, the leading term is of course simply $\chi_1$, of length $1$. The first-order term $v_1 \propto {\cal G}_2{\cal H}_1v_0$ contains terms of length 3, (i.e.\ including $\chi_2$, i.e.\ $p_3\ne 0$), as can be seen explicitly in the $n=3$ case in (\ref{Psifirst3}). The second-order term $v_2$ is found by inverting ${\cal J}$ acting on (\ref{nextit}). The important thing to note is that the leading terms in (\ref{nextit}), those of length 4, are not contained within ${\cal H}_1v_1$, but rather are contained entirely within
\[ {\cal H}_3v_1\propto {\cal H}_3{\cal G}_2 {\cal H}_1v_0\ .\]
The reason is that all pieces of $v_1$ have at most length 3, while ${\cal H}_1$ acts non-trivially only on the indices $p_1,p_2$. Thus all pieces of ${\cal H}_1v_1$ still have at most length 3;  only ${\cal H}_3v_1$ can have pieces with length 4. In fact, since $p_4=0$ in all pieces of $v_1$, all pieces of ${\cal H}_3v_1$ {\em must} have $p_4\ne 0$. This means that ${\cal H}_4$ can be inverted when acting on ${\cal H}_3v_1$, because $p_4\ne 0$ while $p_5=0$. Although the resulting state
\[ {\cal G}_4 {\cal H}_3 {\cal G}_2 {\cal H}_1 v_0 \]
is only a part of $v_2$, all the leading terms in $v_2$ are contained within it, because the only way to get a state of length 5 in $v_2$ at this order is for ${\cal G}_4$ to act on a state with length 4. Repeating this argument at each order means that the expression
\be
\Psi_{\rm leading} = v_0 - \frac{f_1}{J_1}\, {\cal G}_2{\cal H}_1 v_0 + \frac{f_1f_2}{J_1J_2}\, {\cal G}_4 {\cal H}_3 {\cal G}_2 {\cal H}_1 v_0 - \frac{f_1f_2f_3}{J_1J_2J_3} \,{\cal G}_6{\cal H}_5{\cal G}_4 {\cal H}_3 {\cal G}_2 {\cal H}_1 v_0 + \dots
\label{Psileading}
\ee
contains {\em all} the leading terms in $\Psi$.

This shows that whenever acting with ${\cal H}_{2l+1}$ on the leading terms in $v_l$, the resulting state has length $2l+2$ and that ${\cal G}_{2l+2}$ inverts ${\cal H}_{2l+2}$ properly; no corrections here are necessary. Corrections can arise only from subleading terms, i.e.\ those coming from acting on $v_l$ with ${\cal H}_{2j-1}$ where $j\le l$. Since terms in $v_l$ obtained from $\Psi_{\rm leading}$ are at most of length $2l+1$, the only states $|0_a\rangle\rangle$ that can have
\[
\langle\langle 0_a | {\cal H}_{2j-1} |v_l \rangle\rangle \ne 0\ , \qquad j\le l\ ,
\]
 necessarily have length at most $2l+1$. Thus any resulting corrections necessarily have length at most $2l+1$, and so the corrected $v_l$ does not increase in length. The iteration procedure works!
 
The only remaining question is the radius of convergence of the expansion, i.e.\ the largest value of $f/J$ such that $\Psi^\dagger\Psi$ is finite as $L\to\infty$. As shown in the previous section, for $n=3$, the expansion parameter seems to be $f/(2J\sin(3\phih))$. The expansion parameter for general $n$ then seems to be $f/(JD)$, where $D$ is the smallest value of ${\rm det}\,\JQ = \prod_{r=0}^{n-1} \lambda^{(Q,r)}$ for the given $\ah_m$. Making $f<JD$ however is not enough to demonstrate convergence, since at each order in $f/J$ there are multiple terms, as is obvious from (\ref{Psifirst3}). The number of such terms is exponentially growing, so conceivably the radius of convergence is smaller; it is not possible to know exactly how much smaller without more explicit information about $\Psi$. Nonetheless, for sufficiently small $f/J$, the series will converge. An interesting open question is if the convergence of this series is related to the phase diagram of the chain, which is much more complicated than in the Ising case \cite{Huse,Perkoverview}.

\section{Conclusion}
\label{sec:conclusion}

I have shown that there exist exact parafermionic edge zero modes in $\zn$-invariant spin systems for sufficiently small $f/J$, as long as the interactions are chiral. This provides a natural generalization of the results for the Ising/Majorana chain, and indicate that these simple-to-define (but not to analze!) systems possess topological order. It would be very interesting to understand how these fit into the classification schemes of \cite{ChenGuWen,Turner,Schuch}. 

A seeming major difference between the systems studied here and the Ising-Majorana chain is that while 1d fermions exist in nature by trapping electrons in one dimension, it is hard to imagine a parafermion chain being formed in the same fashion. Nonetheless, 1d parafermionic physics can occur in edges of the Read-Rezayi states for the fractional quantum Hall effect \cite{RR}. The wave functions themselves are built from the correlation functions of conformal field theory \cite{FZ2} describing the critical point of the ferromagnetic chain \cite{FZ1}. In fact, the correlators are those of the parafermions themselves; in the $\ztwo$ Moore-Read case \cite{MR}, these are those of Majorana fermions, given by a Pfaffian.
While these Hall edges are purely chiral (i.e.\ have excitations moving in only one direction) and are gapless, gapped parafermionic systems resembling the ones considered here can be made by coupling two different Hall edges together \cite{Clarke,Lindner}. Similar behavior also arises from coupled quantum wires \cite{TeoKane}, and by including defects in multi-component Hall systems \cite{BarkQi}. Thus it is not at all impossible for the topological order described here to occur in a real system.

One particularly interesting more formal direction to explore is to understand what happens for closed boundary conditions. Obviously, there can be no edge modes when there are no edges, but there is still topological order. In the Ising/Majorana case, topological order be detected by computing the sign of the Pfaffian of a matrix determined by the couplings $f_j$ and $J_j$. This is a one-dimensional analog of computing the Chern number to probe topological order in two-dimensional free-fermion systems. No analog of this is yet known for parafermions in any dimension. Is it possible that there is a ``Read-Rezayian'' generalizing the Pfaffian here as well as in the fractional quantum Hall case?

Another direction worthy of further exploration is to understand if similar $\zn$-invariant spin models in higher dimensions exhibit topological order. The Kitaev honeycomb model \cite{Kithoneycomb} in two dimensions arises naturally from coupling together $\ztwo$ spin chains so that the model equivalent to free 2d fermions in a background $\ztwo$ gauge field. Topological order results upon breaking of time-reversal symmetry. It is not difficult to see how to couple together $\zn$-invariant chains to give a model with a background $\zn$ gauge field (an observation also made independently by X.\ Qi, and by C.\ Kane). However, since there is no such thing as a free parafermion in 1d, there is not likely a notion in 2d either. Thus analysis of the resulting model is difficult, but perhaps the exact 1d zero modes found here will be useful in 2d as well. At minimum, a strong hint from the results here is that it would be a good idea to make the interactions chiral by including these phases in the nearest-neighbor terms, a very natural option here.

Finally, it would also be interesting to reverse the arrow of logic and understand what thinking about topological order can say about the original spin system. A result I will discuss in a separate paper is that the integrable couplings (\ref{alphaphi},\ref{fJphi}) can be found simply by demanding that for closed boundary conditions, there exist a shift mode,  generalizing the zero mode found here for open boundary conditions. This gives a direct and simple proof of the observations of \cite{Howes,Gehlen} about the similarity in spectrum between sectors. Many other insights about the integrable spin systems using parafermions also appear possible.

\bigskip

{\em Acknowledgments:} This work was initiated in 2010 and continued while I was visiting Microsoft Station Q in 2010-2011; I am grateful to Michael Freedman and the entire group for their hospitality and generosity. I am also grateful to Eduardo Fradkin, Andrei Bernevig and Taylor Hughes for many interesting discussions, and to Jason Alicea, David Clarke and Kirill Shtengel for discussions of their work. My research is supported by the National Science Foundation under the grant DMR/MPS1006549.

\appendix

\section{Many other parafermionic zero modes}

As mentioned in section (\ref{sec:proof}), the ``Hamiltonian'' ${\cal H}$ has at least $n^L$ zero eigenvalues. Thus the original Hamiltonian has at least $n^L$ zero modes comprised of parafermions, albeit with no guarantee that these remain normalizable as $L\to\infty$. As also mentioned, ${\cal H}$, acting on a $n^{2L}$-dimensional Hilbert space, is equivalent to the sum of two Hamiltonians, each equivalent (up to an overall sign) to the original Hamiltonian $H_n$. The purpose of this appendix is to provide a proof of these claims.

The matrix elements of ${\cal H}$ are given in (\ref{calHdef}) and (\ref{calHdef2}). Ignoring the expression in parentheses, the matrix elements are exactly the same as those in $H_n$. This suggests splitting ${\cal H}$ into two pieces ${\cal H}= \Hone + \Htwo$, arising respectively from the first and second terms in the parentheses. (This is {\em not} the same as splitting it into flip and potential terms ${\cal F}$ and ${\cal J}$.) Doing this shows that not only are $\Hone$ and $-\Htwo$ each equivalent to the original Hamiltonian, they commute with each other.  This allows a simple proof of the existence of exact zero modes; each corresponds to an eigenvector of $\Hone$ and $\Htwo$ with opposite eigenvalues of the two.

Since the ``Hamiltonian'' ${\cal H}$ acts on the states $|p_1p_2\dots\rangle\rangle$ by simply shifting the $p_i$, with matrix elements depending only on the values $p_i$, it can be rewritten in terms of operators acting as (\ref{explicitrep}). They thus obeying the exact same algebra (\ref{znspin1},\ref{znspin2}) as $\sigma_j$ and $\tau_j$ do, only now acting on the new $2L$-``site'' Hilbert space. To avoid confusion, I call the diagonal and shift operators $s_a$ and $t_a$ respectively.
Using (\ref{calHdef},\ref{calHdef2}) then gives
\begin{eqnarray}
\Hone &=& -\sum_{a=1}^{2L-1} u_a \sum_{m=1}^{n-1}
\omega^{m(m-n)/2}\alpha_m 
 (t_a^\dagger t^{}_{a+1})^m s_a^{-m}\ ,\\
 \label{H1new}
\Htwo &=& \sum_{a=1}^{2L-1} u_a \sum_{m=1}^{n-1}
\omega^{m(m-n)/2}\alpha_m 
 (t_a^\dagger t^{}_{a+1})^m s_{a+1}^{m}\ ,
\label{H2new}
\end{eqnarray}
where $u_{2j-1}= f_j$ and $u_{2j}= J_{j}$ and for simplicity here I set $\alpha_m=\ah_m$.
These two Hamiltonians indeed commute term by term.
These new Hamiltonians are quite simple in the Ising/Majorana case. Since $s_a$ and $t_a$ here are the Pauli matrices $\sigma^z_a$ and $\sigma^x_a$, using $\sigma^y=i\sigma^x \sigma^z$ gives
\[
\Hone = -\sum_{a=1}^{2L-1} u_a \,\sigma^y_a \sigma^x_{a+1}\ , \qquad 
\Htwo = \sum_{a=1}^{2L-1} u_a \,\sigma^x_a \sigma^y_{a+1}\ .
\]
The equivalence of the two Hamiltonians is obvious, coming from a canonical transformation exchanging $\sigma^x_a$ and $\sigma^y_a$ at each site, sending $\Hone\leftrightarrow -\Htwo$. Another amusing thing to note is that if the canonical transformation exchanging $\sigma^x_a$ and $\sigma^y_a$ is instead done at every other sites ${\cal H}$ simply becomes the famed quantum XY Hamiltonian. Splitting it into $\Hone$ and $\Htwo$ then amounts in fermion language to splitting a free complex fermion into two Majorana theories.

To see the equivalence between $\Hone$ and $-\Htwo$ and the original Hamiltonian, first note 
that all of the terms in $\Hone$ satisfy the same algebra as do the terms in the original Hamiltonian. For example,
\[(t_a^\dagger t^{}_{a+1} s^\dagger_a) (t_{a+1}^\dagger t^{}_{a+2} s^\dagger_{a+1}) = \omega\,
 (t_{a+1}^\dagger t^{}_{a+2} s^\dagger_{a+1})  (t_a^\dagger t^{}_{a+1} s^\dagger_a)\ ,\]
 just like $\tau_a (\sigma_a^\dagger\sigma_{a+1}) = \omega (\sigma_a^\dagger\sigma_{a+1}) \tau_a$.
 The same is true for $\Htwo$. Moreover, the $n^{2L}$-dimensional Hilbert space can be split into two $n^{L}$-dimensional spaces, one of which is found by acting with the terms in $\Hone$ on some reference state (say $|00\dots\rangle\rangle$), and the other by acting with $\Htwo$ on the same reference state. Thus within each of these Hilbert spaces, $\Hone$ and $\Htwo$ are equivalent to the original Hamiltonian $H_n$, up to the sign in front of $\Htwo$. 
 
This equivalence requires $\Hone$ have the same spectrum as $H_n$, while $\Htwo$ has the opposite spectrum. Since the two can be simultaneously diagonalized, ${\cal H}=\Hone + \Htwo$ must have at least $n^{L}$ states with eigenvalue zero. Thus the original problem has many zero modes comprised of parafermions!  The harder part is now showing that one of these zero modes is a normalizable edge zero mode. This is the proof of section \ref{sec:proof}.

\end{document}